\documentclass[11pt]{article}

\usepackage{pb-diagram}
\usepackage{latexsym}
\usepackage{amsfonts}
\usepackage[usenames,dvipsnames]{xcolor}
\usepackage{graphicx,amssymb,amsmath,epsfig}
\usepackage[english]{babel}
\usepackage{graphicx}
\usepackage{dcolumn}
\usepackage{bm}

\definecolor{myblue}{rgb}{0,0,0.8}
\usepackage[colorlinks=true
,urlcolor=myblue	
,anchorcolor=myblue
,citecolor=myblue
,filecolor=myblue
,linkcolor=myblue
,menucolor=myblue
,pagecolor=myblue
,linktocpage=true  
]{hyperref} 

\definecolor{green}{RGB}{0, 130, 0}
\definecolor{grey}{RGB}{90, 90, 90}

\catcode`\@=11
\def\marginnote#1{}

\newcount\hour
\newcount\minute
\newtoks\amorpm
\hour=\time\divide\hour by60 \minute=\time{\multiply\hour by60
\global\advance\minute by-\hour}\edef\standardtime{{\ifnum\hour<12
\global\amorpm={am}%
        \else\global\amorpm={pm}\advance\hour by-12 \fi
        \ifnum\hour=0 \hour=12 \fi
        \number\hour:\ifnum\minute<10
0\fi\number\minute\the\amorpm}}
\edef\militarytime{\number\hour:\ifnum\minute<10 0\fi\number\minute}

\def\draftlabel#1{{\@bsphack\if@filesw {\let\thepage\relax
   \xdef\@gtempa{\write\@auxout{\string
      \newlabel{#1}{{\@currentlabel}{\thepage}}}}}\@gtempa
   \if@nobreak \ifvmode\nobreak\fi\fi\fi\@esphack}
        \gdef\@eqnlabel{#1}}
\def\@eqnlabel{}
\def\@vacuum{}
\def\draftmarginnote#1{\marginpar{\raggedright\scriptsize\tt#1}}
\def\draft{\oddsidemargin -.5truein
        \def\@oddfoot{\sl preliminary draft \hfil
        \rm\thepage\hfil\sl\today\quad\militarytime}
        \let\@evenfoot\@oddfoot \overfullrule 3pt
        \let\label=\draftlabel
        \let\marginnote=\draftmarginnote

\def\@eqnnum{(\theequation)\rlap{\kern\marginparsep\tt\@eqnlabel}%
\global\let\@eqnlabel\@vacuum}  }

\def\numberbysection{\@addtoreset{equation}{section}
        \def\theequation{\thesection.\arabic{equation}}}

\def\underline#1{\relax\ifmmode\@@underline#1\else
 $\@@underline{\hbox{#1}}$\relax\fi}

\catcode`@=12 \relax

\numberbysection

\topmargin by -\headsep
\def\nonu{\nonumber}
\def\br{\begin{eqnarray}}
\def\er{\end{eqnarray}}
\def\lb{\lbrack}
\def\rb{\rbrack}

\def\({\left(}
\def\){\right)}
\def\[{\left[}
\def\]{\right]}

\def\lie{{\cal G}}
\def\a{\alpha}
\def\b{\beta}

\def\d{\delta}

\def\eps{\epsilon}

\def\l{\lambda}

\def\pa{\partial}

\def\s{\sigma}

\def\tp0{\Theta_{+}^{(0)}}
\def\tm0{\Theta_{-}^{(0)}}

\def\cm{{\cal M}}

\def\l{\lambda}

\def\nonu{\nonumber}

\def\bi{\begin{itemize}}
\def\ei{\end{itemize}}

\def\ck{{\cal K}}

\newcommand{\purple}[1]{{\color{purple}#1}}

\begin{document}

\today 
\hskip 5cm generalized miura-jfg-revised.tex
\vspace*{1cm}
\noindent

\begin{center}
{\Large\bf     Gauge Miura and B\"acklund Transformations for Generalized $A_n$-KdV  Hierarchies}
\end{center}
\normalsize
\vskip 1cm
\begin{center}
{J.M. de Carvalho Ferreira}\footnote{\href{mailto:jogean.carvalho@unesp.br}{jogean.carvalho@unesp.br}},   J.F. Gomes\footnote{\href{mailto:francisco.gomes@unesp.br}{francisco.gomes@unesp.br}}, G.V. Lobo\footnote{\href{mailto:gabriel.lobo@unesp.br}{gabriel.lobo@unesp.br}}, and A.H. Zimerman\footnote{\href{mailto:a.zimerman@unesp.br}{a.zimerman@unesp.br}}\\[.7cm]

\par \vskip .1in \noindent
 {Instituto de F\'isica Te\'orica - IFT/UNESP,\\
Rua Dr. Bento Teobaldo Ferraz, 271, Bloco II,
CEP 01140-070,\\ S\~ao Paulo - SP, Brasil.}\\[0.3cm]

\vskip 2cm

\end{center}

\begin{abstract}

The construction of  Miura  and B\"acklund transformations for $A_n$ mKdV and KdV hierarchies  are presented in terms of gauge transformations  acting 
 upon the zero curvature representation.  As in the well known  $sl(2)$ case, we  derive and relate the equations of motion  for the two hierarchies. Moreover, the Miura-gauge  transformation is not unique, instead, it is shown to be connected  to a set of generators  labeled by the exponents  of $A_n$
The construction of generalized gauge-B\"acklund transformation  for the $A_n$-KdV hierarchy is obtained as a composition of Miura and B\"acklund-gauge transformations  for $A_n$-mKdV hierarchy. The zero curvature representation provide  a framework  which is universal within all flows    and  generate systematically  B\"acklund  transformations for the  entirely hierarchy.
\end{abstract}

\vskip 1 cm
\section{Introduction}

The recent  studies  of integrable models has revealed  an increasing  importance of  its underlying algebraic structure,  which is intimately  connected to a series of peculiar properties,  such as  the existence of an infinite number of conservation laws,
 soliton solutions,  B\"acklund transformation, etc. Structural connection  has been known  for some time, see for instance,  \cite{drinfeld-sokolov}, \cite{miramontes}, \cite{burroughs}.
 Many of these properties  can be derived from the  zero curvature formulation in terms of two dimensional gauge potentials, $A_{x}$ and $A_{t}$  lying  in an affine algebra, $\hat \lie$, i.e.,
 \br
 \pa_xA_t-\pa_tA_x +[A_x,A_t]=0. \label{zcc1}
 \er
 A particular virtue of the zero curvature representation (\ref{zcc1}) is  its  gauge invariance.
 Conservation laws  can be derived  by  abelianizing  the two dimensional   potentials  by  gauge transforming them into the  Cartan subalgebra of $\hat \lie$  \cite{olive-turok}.  Soliton solutions in turn,   can be constructed  by gauge transforming   a vacuum solution $A_x^{vac}$ and $A_{t}^{vac}$  into  some nontrivial configuration (dressing) see for instance \cite{babelon-book}.  More recently, gauge transformation     was shown to  be  a key ingredient in  constructing   B\"acklund transformations \cite{thiago}-\cite{lobo}.
 In all cases, gauge transformation  is an essential  ingredient.

   
  { {
In this paper we  shall consider the multicomponent  mKdV  and KdV  hierarchies   within the generalized Drinfeld-Sokolov matrix hierarchy
  connected to  the affine 
$\lie =A_{n}$ algebra discussed  in \cite{drinfeld-sokolov}.
Also important are its connection to
several physically relevant nonlinear soliton equations,    $W$ algebras  and discrete matrix models  \cite{miramontes},\cite{burroughs}, \cite{bakas1}, \cite{bakas2}.}}


  B\"acklund transformation interpolates between two soliton solutions of an integrable  model \cite{rogers}.  More recently,
 it was employed  to describe  integrable defects in the sense that  two field configurations  can be connected  
 by a defect at certain space location \cite{corrigan1}-\cite{corrigan}.  Thus, the classification of integrable  defects and  the construction of the various types of B\"acklund transformations are intimately connected.

 The construction of B\"acklund transformations can be formulated  in terms of gauge transforming the two dimensional  potentials  $A_{\mu}$  from    one  field configuration, say $\phi$  to another,  $\psi$, i.e.,
 \br
U(\phi_i,\psi_i) A _{\mu}(\phi_i) = A _{\mu}(\psi_i)U(\phi_i,\psi_i)  +\pa_{\mu}U(\phi_i,\psi_i) 
\label{gauge}
\er
 A natural  construction  of  B\"acklund-gauge transformations $ U(\phi_i,\psi_i) $
 in terms of  a graded   affine algebra  was proposed  in  \cite{ana1}, \cite{ana-miura} for  the $A_1$ mKdV (sinh-Gordon)   and in  \cite{lobo} for the generalized $A_n$ Toda  Hierarchies. 
 
  Moreover   $U(\phi_i,\psi_i)$  was    shown  to be an universal object  and provide a systematic construction of B\"acklund transformation extended  to   all evolution equations (flows) within the hierarchy  { { since  (\ref{gauge}) is valid for all flows (i.e. $A_t \equiv A_{t_N}$).}}
 
 Another application of gauge invariance of the zero curvature formulation  is  the construction of a Miura transformation connecting  the   mKdV and KdV hierarchies.   
 {  {Results in this direction were obtained for the  $A_1$ loop algebra involving a single field  in ref. \cite{d-sok}, \cite{manas}.  In this paper we shall discuss the construction of the Miura transformation  for the  multicomponent 
 $A_n$  affine algebra.  We shall   also discuss the interrelation   between the  B\"acklund transformation of  the two systems.}}
 Several aspects  of  KdV hierarchies  have  been developed  in the past with regards to  spherically symmetric   reductions of self dual Yang-Mills theories  in 4-D,  the realization of $W$-algebras, etc (e.g.  see for instance \cite{fordy1}-\cite{mathieu} and refs. therein). 
In particular, in ref. \cite{ana-miura}  a Miura-gauge transformation $S$  was proposed to map   for  $sl(2)$ the   two dimensional  potentials $A_{\mu}^{mKdV} $ into $A_{\mu}^{KdV}$,  i.e.,
 \br
 A_{\mu}^{KdV} = S A_{\mu}^{mKdV} S^{-1} + S \pa_{\mu} S^{-1}\label{miura-gauge12}
 \er

  In this paper we follow  the same line of reasoning of ref. \cite{ana-miura}  to discuss the construction of Miura-gauge transformation  from the  $A_n$-mKdV to the  $A_n$-KdV systems.  {  { An important consequence  of our construction in (\ref{gauge})  is that it is valid for  all flows  $t=t_{N}$ and henceforth  (\ref{gauge})   provide a  systematic  derivation of B\"acklund-gauge  transformation  for all flows of  the $A_n$-KdV  from the $A_n$-mKdV system  as a composition of $ U$ and $S$(see eq. (\ref{b-g-kdv1})). }}


This paper is organized as follows.
 In sections 2  and 3 we  discuss the  algebraic structure and construct  the general  $A_n$  equations of motion  for the second flow  of zero curvature  representation for both mKdV and KdV systems.  In section 4  we discuss the 
  structure of the Miura-gauge  transformation for $A_1, A_2 $ systems (and for $ A_3$  in the appendix B).  In particular, we  argue that there are  more than one  Miura transformation connecting both  systems.  In fact, for $A_n$ there are  $n+1$  transformations labeled by the identity $I$ and  by the discrete set of 
  constant  generators  $E^{(k)} \in  \hat \ck$,  of grade $k=1, \ldots, n$. We construct explicitly the  Miura transformations for $A_1, A_2$ and $A_3$  and  propose a general  construction for the $A_n$ case.    Section 5  contains     the general  structure  between equations of motion of the two integrable hierarchies.  In section 6 we present the  construction of    B\"acklund transformation  for the $A_n$-KdV system as a combination of  Miura and B\"acklund transformations for the mKdV system.  Section 7  contains  our conclusions.

\section{mKdV Equations}

Here we  shall 
discuss in detail the general construction of the $A_n$-mKdV flows. {   { We shall use $``n''$ for the rank  of the underlying  $\lie = A_n$ affine algebra  (which also  equals  the number of fields of the theory) and     $``N''$, to label the flows (or time evolution according to $t_N$).   }}

Consider the generic time evolution equations  for the  $A_n$-mKdV hierarchy which are classified according  to  the grading structure developed in the Appendix A. 
Consider the decomposition of an affine Lie algebra $\hat \lie = \sum_a \lie_a$  according to  a grading operator $Q$ such that $[Q, \lie_a]=a\lie_a$ and $[ \lie_a ,\lie_b]\subset \lie_{a+b}$.
Let    $E \equiv E^{(1)}$  be a constant grade one generator  responsible for a second decomposition  of 
  $\hat \lie$ into  Kernel  of $E$, $\hat \ck =  \{ x \in  \hat \ck, [x, E]=0\}$  and its complement, $\hat \cm$ i.e.,
\br
\Hat \lie =\hat { \ck} \oplus \hat {\cm}.
\er
In particular, projecting into the zero grade subspace, $\lie_0 = \ck\oplus \cm  $.   Define  now  the Lax operator as
\br
L=\pa_x + E+A_0,
\label{lax}
\er
where $A_0  \in \cm$ (and consequently $A_0 \in \lie_0$).   In the case of $\hat \lie = \hat {sl}(n+1)$ and  {\it principal gradation}   with subspaces given as in (\ref{a4}), 
$E^{(1)}$  is given by
\begin{equation}
E \equiv E^{(1)} = \sum_{k=1}^{n} E_{\alpha_k} + \lambda
E_{-(\alpha_1+{\cdots} +\alpha_{n})} \, .
\label{Eone}
\end{equation}
and
$A_0 $  parameterized as
 \br 
 A_0= \sum_{k=1}^{n} v_k h_{k} = \left( \begin{matrix} r_1&0&\cdots
&0 \\0&r_2 &\cdots&0\\
 \vdots&\vdots &\ddots& \vdots\\ 
 0&0 &\cdots &r_{n+1} \end{matrix}
\right) \, ,\label{a0}
 \er
 where $r_i = v_i-v_{i-1}, \;\;i=1,{\ldots} , n+1, \;\;\; v_0 = v_{n+1}=0$.  The relation $v_i =\sum_{j=1}^i r_j$
follows from trace  condition.

We now propose  the construction of  time evolution equations in the zero curvature representation,
\br
\lb \pa_x+ A_x, \pa_{t_N} + A_{t_N}\rb =\pa_xA_{t_N}- \pa_{t_{N}}A_x  +\lb A_x, A_{t_N} \rb = 0
\label{zcc}
\er
where $A_x = E^{(1)}+ A_0$.

 The  so called {\it  positive grade}  time evolution   {  {according to $t_N$ is  constructed  from   $A_{t_N}$    decomposed as }}
	\br
	A_{t_N} = D^{(N)} + D^{(N-1)} {+} \cdots +D^{(0)}, \qquad D^{(j)} \in \lie_j, \quad N\in {\mathbb{Z}}_{+}.
	\label{atn}
	\er
	The zero curvature representation (\ref{zcc})  decomposes  according  to the graded structure  into
	\br
	[E^{(1)}, D^{(N)} ]&=&0,\label{1} \\
	\lb E^{(1)}, D^{(N-1)} \rb + \lb A_0, D^{(N)}\rb +\pa_x D^{(N)} &=&0,\label{2} \\
	 & \vdots  & \nonu \\
	\lb A_0, D^{(0)}\rb  +\pa_x D^{(0)} - \pa_{t_N} A_0 &=&0, \label{a5}
	\er 
	and allows  solving for $D^{(j)}$ recursively starting from the highest grade  eqn. (\ref{1}).   In particular, the last eqn. (\ref{a5})  is the only eqn. involving  time 
	derivatives acting on  $A_0$ and can be  regarded as the time evolution for the  fields  parametrizing  $A_0 \in \cm$.  
	
		Solving (\ref{1})  for $N=2$ and $n>1$, according to the grading structure given in (\ref{a4}) we find
		
		\begin{equation}
D^{(2)}= a_2\( \sum_{k=1}^{n-1} E_{\alpha_k+\alpha_{k+1}}
+ \lambda E_{-(\alpha_1+\cdots+\alpha_{n-1})}+
\lambda E_{-(\alpha_2+\cdots+\alpha_{n})}\)  \in \hat \ck \, .
\label{D2}
\end{equation}
 where $a_2$ is an arbitrary  coefficient.  The lower grade matrices $D^{(i)}, i=0,1$ can be solved recursively
from the appropriate grade projections of the zero-curvature equation 
(\ref{zcc}.) For example when $N=2$, the grade one element $D^{(1)}$ is solved from (\ref{2})  i.e.,
\br
\lbrack  E^{(1)}, D^{(1)} \rbrack + \lbrack A_0, 
D^{(2)}\rbrack  +\partial_x D^{(2)} =0 \, .
\er
 leading  to
\begin{equation}
 \pa_x a_2 = 0 \qquad {\rm and} \qquad D^{(1)}= \sum_{k=1}^{n} (v_{k+1}-v_{k-1})\,  E_{\alpha_k} + (v_1-v_n )\lambda
 E_{-(\alpha_1+{\cdots} +\alpha_{n})}
 \label{Done}
 \end{equation}
 {where} we shall set  $a_2=1$.
%

The grade $1$ projection of the zero-curvature equation(\ref{zcc}) leads to
\br
\lbrack E^{(1)}, D^{(0)} \rbrack  + \lbrack A_0, D^{(1)}\rbrack  
+\partial_x D^{(1)} =0,
\er
with $D^{(0)}= \sum_{a=1}^{n} d_a h_{a} $.  It therefore follows,
\br
{\mathbf K}_{ja} d_a= {\mathbf K}_{ja}v_a(v_{j+1}-v_{j-1}) + \pa_x (v_{j+1}-v_{j-1}), \label{frank1}
\er
where ${\mathbf K}_{ab} = 2{{\a_a\cdot \a_b}\over {\a_b^2}} = 2\d_{a,b}-\d_{a,b-1}-\d_{a,b+1}$ is the Cartan matrix of $sl(n+1)$.
%
%
%
%
%

%

Solving \eqref{frank1} for $d_a$, $a=1,\ldots,n$ and inserting it into the grade $0$ projection of 
the zero-curvature equation \eqref{a5} yields the eqn. of motion 
for 
$t \equiv t_2$,
\begin{equation}
\partial_t { v_a}  = \partial_x { d_a}\, . %
\label{vectorrel}
\end{equation}
After multiplying  both sides of relation \eqref{vectorrel}
by the Cartan matrix ${\mathbf K}$ and making use of relation \eqref{frank1} 
we obtain,  \cite{alves1}

\begin{equation}
\begin{split}
\partial_t \sum_{{a}=1}^{n}  {\mathbf K}_{ba} v_{a}&=
\partial_t \left( 2 v_b - v_{b+1}-v_{b-1}\right) \\
&=  \partial_x  \bigg\lbrack \partial_x (v_{b+1}-v_{b-1}) + (2v_b-v_{b+1}-v_{b-1})
(v_{b+1}-v_{b-1}) \bigg\rbrack, \;\; b=1, {\ldots} , n \, .
\end{split}
\label{frank2}
\end{equation}
Recalling 
equation \eqref{frank2} can be rewritten as,\begin{equation}
\partial_t   \left( r_{b+1}-r_{b} \right)=
- \partial_x  \bigg\lbrack \partial_x ( r_b+r_{b+1} ) - \left( r_{b+1}-r_{b} \right)
( r_b +r_{b+1}) \bigg\rbrack, \quad b=1, {\ldots} , n\, ,
\label{5prime}
\end{equation}
that defines $t_2$ flows of $sl(n+1)$ mKdV hierarchy.  
 Inverting the Cartan matrix explicitly for the  cases where $n=2,3 $   in (\ref{frank2}) we find  in terms of $v_j$ variables, respectively,
 \begin{equation}
	\begin{split}
	\pa_{t_2} v_1 =& {{1}\over {3}}\pa_x\(-\pa_xv_1+2\pa_xv_2+v_1^2 - 2v_2^2 + 2v_1v_2\),  \\
	\pa_{t_2} v_2 =& {{1}\over {3}}\pa_x\(-2\pa_xv_1+\pa_x v_2+2v_1^2 - v_2^2 - 2v_1v_2\), \\
	\label{eqmot3}
	\end{split}
	\end{equation}
	and 
	\begin{equation}
	\begin{split}
	\pa_{t_2} v_1 =& {{1}\over {2}}\pa_x\(-\pa_xv_1+\pa_xv_2+\pa_xv_3+v_1^2 - v_2^2 -v_3^2+ v_1v_2 +v_2v_3 \),  \\
	\pa_{t_2} v_2 =& \pa_x\(-\pa_xv_1+\pa_xv_3 +v_1^2 -v_3^2 - v_1v_2+v_2v_3\), \\
	\pa_{t_2} v_3 =& {{1}\over {2}}\pa_x\(-\pa_xv_1-\pa_x v_2+\pa_xv_3+v_1^2 +v_2^2 -v_3^2- v_1v_2-v_2v_3\), \\
	\label{eqmot4}
	\end{split}
	\end{equation}
\section{KdV Equations}

For the $A_n$-KdV hierarchy we  shall employ the same  algebraic structure namely, principal gradation, $Q^{ppal}$   and constant grade one  element, $E^{(1)}$ (\ref{Eone})  and propose the  following  Lax operator 
\br
L=\pa_x + E^{(1)} +A^{(-1)}+\cdots + A^{(-n)} , \qquad A^{(k-n-1)}= J_{n+1-k} E_{-(\a_k + \cdots + \a_n)} \in \hat {\cal {M}} \subset \lie_{k-n-1}
\label{kdv-lax}
\er
for $k=1, \ldots ,n$, or in matrix form
 \br
 A_x =  E^{(1)} +A^{(-1)}+\cdots + A^{(-n)}=\left( \begin{matrix} 0&1&0&\cdots &0&0 \\ 0&0 &1&\cdots&0&0\\ & \ddots
&&\ddots& &\vdots\\
& &  \ddots & &1&0\\ 
 \vdots&& &  &0&1\\ 
 \l+J_n&J_{n-1} &\cdots & 0&J_1&0 \end{matrix}
\right)
\er
  The zero curvature representation  {for    flow $t= t_2$,  (i.e., $N=2$), i.e., }
\br
\lb \pa_x + E^{(1)} + A^{(-1)} +\cdots + A^{(-n)},  \;\;\pa_{t_2} + D^{(2)}+D^{(1)} + \cdots + D^{(-n)}\rb =0
\label{zcc-kdv}
\er
  leads to the following graded eqns.,
\br
&&\lb E^{(1)}, D^{(2)} \rb=0, \label{d2}\\
&&\lb E^{(1)}, D^{(1)} \rb+\pa_xD^{(2)}=0, \label{d1}\\
&&\lb E^{(1)}, D^{(0)} \rb + \lb A^{(-1)}, D^{(2)}\rb +\pa_xD^{(1)}=0, \label{d0} \\
&&\lb E^{(1)}, D^{(-1)} \rb + \lb A^{(-1)}, D^{(1)}\rb + \lb A^{(-2)}, D^{(2)}\rb +\pa_xD^{(0)}=0, \label{d-1}
\er
which involves the unknowns  $D^{(2)}, D^{(1)}, D^{(0)}$ and $D^{(-1)}_{{\cal M}}$. 
Moreover, the lower  graded   eqns.,
\br
 \lb A^{(-1)}, D^{(-n)}\rb + \ \lb A^{(-2)}, D^{(-n+1)}\rb + \cdots  + \lb A^{(-n)}, D^{(-1)}\rb &=& 0, \label{d-1k} \\
  \lb A^{(-2)}, D^{(-n)}\rb + \ \lb A^{(-3)}, D^{(-n+1)}\rb  + \cdots  + \lb A^{(-n)}, D^{(-2)}\rb &=&  0, \\
&\vdots &\nonu \\
  \lb A^{(-n)}, D^{(-n+1)}\rb + \ \lb A^{(-n+1)}, D^{(-n)}\rb &=& 0,\\
    \lb A^{(-n)}, D^{(-n)}\rb &=& 0 \label{d-n}.
    \er
  involves   the unknowns  $D^{(-1)}_{{\cal K}},  \cdots, D^{(-n)}$. 
 Together  they  lead to the time evolution eqns.
\br
\pa_{t_2}A^{(-1)} &=&\lb E^{(1)}, D^{(-2)} \rb + \lb A^{(-1)}, D^{(0)}\rb + \lb A^{(-2)}, D^{(1)}\rb   + \lb A^{(-3)}, D^{(2)}\rb + \pa_xD^{(-1)}, \label{a-1}\\ 
\pa_{t_2}A^{(-2)} &=& \lb E^{(1)}, D^{(-3)} \rb + \lb A^{(-1)}, D^{(-1)}\rb +  \cdots  + \lb A^{(-4)}, D^{(2)}\rb + \pa_xD^{(-2)}, \label{a-2}\\
 &\vdots & \nonu \\
\pa_{t_2}A^{(-n)} &=&  \lb A^{(-1)}, D^{(-n+1)}\rb + \ \lb A^{(-2)}, D^{(-n+2)}\rb + \cdots  + \lb A^{(-n)}, D^{(0)}\rb +\pa_x D^{(-n)}. \label{a-n}
\er

In order to obtain  the time evolution  eqns.  (\ref{a-1})-(\ref{a-n})  we need to solve eqns. (\ref{d2})-(\ref{d-1})  for $D^{(2)},  D^{(1)}, D^{(0)}, D^{(-1)}_{\cal M}$ and  (\ref{d-1k})-(\ref{d-n}) for $D^{(-1)}_{\cal K}$, $D^{(-2)}$, $\cdots$, $D^{(-n)}$ .
Let us start from (\ref{d2})  which coincides with (\ref{1})  for $N=2$ and  yields  $D^{(2)} = D^{(2)}_{\cal K} \in \cal K$ given by (\ref{D2}).

 Notice  that   since \footnote{In general the subspaces defined by  $E$ is  such that
\br 
\lb  {\cal{K}, \cal{K} }\rb \in \cal{K}, \quad  \lb \cal{K}, \cal{M} \rb \in \cal{M}, \nonu
\er 
and  each of the above eqns.  gives rise to $\cal K $ and $\cal M$ components.} $\lb E^{(1)}, D^{(1)} \rb \in \cal {M}$,  eqn.  (\ref{d1})  imply that $ \pa_x(a_2 )=0,\;\;  D^{(1)}_{\cal M} =0$ and henceforth $D^{(1)} = D^{(1)}_{\cal K} \in \hat {\cal K}$, i.e.,
\br
{{D^{(1)} }}=  a_1\(E_{\a_1}^{(0)} +  E_{\a_2}^{(0)} + \cdots + E_{\a_n}^{(0)} +  E_{-(\a_1+ \cdots + \a_n)}^{(1)} \)
\er
Eqn (\ref{d0}) imply that
\br
\pa_x a_1 =0, \qquad \lb E^{(1)}, D^{(0)}_{\cal M} \rb + \lb A^{(-1)}, D^{(2)}_{\cal K}\rb =0
\er
 and we shall take $a_1=0$ and $a_2=1$.  Inserting  $A^{(-1)}=J_1\; E_{-\a_n}^{(0)}$ and $D^{(0)} =\sum_{i=1}^{n}d_i\; (\mu_i \cdot H^{(0)})$ we find 
\br
D^{(0)} = -J_1 \;\(\mu_{n-1}\cdot H^{(0)}\), \label{dd0}
\er
where $\mu_{n-1}$ is the $(n-1)-$th fundamental weight of $A_n$ ,i.e.,
\br
\mu_{n-1} =\sum_{i=1}^{n-1} \({{2i}\over {n+1}}\) \a_i +\({{n-1}\over {n+1}}\)\a_n.
\er
where $\a_j, j=1, \cdots n$ are simple roots of $A_n$.  Following the same philosophy,  we find from (\ref{d-1})  that
\br
D^{(-1)}_{\cal M} &=& \sum_{i=1}^{n-2} {{2i}\over {n+1}}\pa_xJ_1 \;E_{-\a_i}^{(0)} +\( 2{{n-1}\over {n+1}}\pa_x J_1 {+} J_{2}\) E_{-\a_{n-1}}^{(0)}\nonu \\
&+&\( {{n-1}\over {n+1}}\pa_x J_1 {+} J_{2}\) E_{-\a_{n}}^{(0)}
\nonu \\
\er

In order to solve  eqns. (\ref{d-1k})-(\ref{d-n})
we propose a { general solution }  for $D^{(-i)}, \; i=1, \ldots, n$  by assigning   to them only negative step operators (lower diagonal) of grade $-i$ as follows,
\br
D^{(-n)} &=& b^{(-n)}_1E^{(0)}_{-(\a_1 + \cdots + \a_n)}, \nonu \\
D^{(-n+1)} &=&  b^{(-n+1)}_1E^{(0)}_{-(\a_1+ \cdots + \a_{n-1})} +  b^{(-n+1)}_{2}E^{(0)}_{-(\a_2 + \cdots + \a_{n})}, \nonu \\
 & \vdots & \nonu \\
D^{(-1)} &=&  b^{(-1)}_1E^{(0)}_{-\a_1} +  b^{(-1)}_{2}E^{(0)}_{-\a_2 },+\cdots  + b^{(-1)}_{n}E^{(0)}_{-\a_n }  \nonu \\
\er
where $b_k^{(-i)}$  are   coefficients  to  be determined by eqns. (\ref{a-1})-(\ref{a-n}).

In order to illustrate our formulation consider  the following explicit  examples,
\begin{itemize}
\item $ \lie = sl(3), \;\; n=2$
\br
D^{(0)} &=& - J_1 \({{2}\over {3}}h_1^{(0)}+{{1}\over {3}}h_2^{(0)}\), \nonu \\
D^{(-1)} &=& (J_{2} + {{2}\over {3}}\pa_x J_1) E_{-\a_1} ^{(0)} +  (J_{2} + {{1}\over {3}}\pa_xJ_1)E_{-\a_2}^{(0)}.
\er
Eqns. (\ref{a-1}) and (\ref{a-2}) give rise  to 
\br
D^{(-2)} &=& - ( {{2}\over {3}}\pa_x^2J_1 +\pa_x J_2)E_{-\a_1-\a_2}^{(0)}
\er
and respectively to the following eqns. of motion
\br
\pa_{t_2}J_1 &=& \pa_x^2J_1+2\pa_xJ_2, \label{j2}\\
\pa_{t_2}J_2 &=& -{{1}\over {3}} \(2\pa_x^3 J_1 +3\pa_x^2J_2-2J_{{1}} \pa_xJ_1\). \label{eq-kdv-sl3}
\er

\item $\lie = sl(4), \;\; n=3$
\br
D^{(0)} &=& - J_2 ({{1}\over {2}}h_1^{(0)}+h_2^{(0)} + {{1}\over {2}}h_3^{(0)}), \nonu \\
D^{(-1)} &=&  {{1}\over {2}} \pa_xJ_1 E_{-\a_1}^{(0)} +  (J_2 + \pa_xJ_1 )E_{-\a_2}^{(0)} +(J_2 + {{1}\over {2}}\pa_xJ_1 )E_{-\a_3}^{(0)}. 
\er
Likewise,  (\ref{a-1}) and (\ref{a-2}) yields,
\br
D^{(-2)} &=&\( -{{1}\over {2}} \pa_x^2J_1+J_3\)E_{-\a_1-\a_2}^{(0)} +   \(-{{3}\over {2}}\pa_x^2J_1-\pa_xJ_2+J_3\)E_{-\a_2-\a_3}^{(0)}  , \nonu \\
D^{(-3)} &=&\({{1}\over {2}}\pa_x^3J_1-\pa_xJ_3\) E_{-\a_1-\a_2-\a_3}^{(0)} 
\er
 and 
  the  eqns. of motion
\br
\pa_{t_2}J_1 &=& 2\(\pa_x^2J_1+\pa_xJ_2\), \\
\pa_{t_2}J_2 &=& {} \( -2\pa_x^3 J_1 -\pa_x^2J_2+J_1\pa_xJ_1 +2\pa_xJ_3\),\\
\pa_{t_2}J_3 &=&{ {1}\over {2}}\(\pa_x^4J_1-J_1\pa_x^2J_1-2\pa_x^2 J_3+J_2\pa_xJ_1\).  \label{eq-kdv-sl4}
\er
\end{itemize}


\section{Miura Transformation}

In this section we consider  $\lie =A_n$ and propose a   Miura-gauge transformation  $S$  to connect the  two   gauge potentials $A_{x}^{mKdV}$    and  $A_{x}^{KdV}$  as,
\br
  A_x^{KdV} = S A_x^{mKdV}S^{-1} + S\pa_x S^{-1}\label{miura-gauge1}
  \er
  where
\br 
A_x^{mKdV} = E^{(1)} + \sum_{i=1}^{n} v_i h_i, 
\qquad 
A_x^{KdV} = E^{(1)} + \sum_{j=1}^{n} J_{{n+1-i}} E_{-(\a_i+ \a_{i+1} + \cdots+ \a_n)} .\label{aa}
\er
The first  thing to notice is that both potentials  in (\ref{aa}) are defined according to the {\it  same Lie algebraic structure},  i.e., principal gradation, $Q^{ppal}$   and share the same  constant, grade one  semi-simple element $E^{(1)}$ given in (\ref{Eone}) (see app. A).  The   desired Miura-gauge transformation  $S$  is then constructed to  preserve such structure.  
Let us recall  the $sl(2)$ case where the Miura  transformation $ J  = \eps \pa_xv-v^2,   \;\; \eps = \pm 1$  connects {{   { { the  two dimensional gauge potentials 
\br
A_x^{mKdV} =\begin{pmatrix}
v & 1 \\
\l & -v 
\end{pmatrix}
\qquad  \quad
A_x^{KdV} =\begin{pmatrix}
0 & 1 \\
\l+J & 0
\end{pmatrix} \label{axsl2}
\er
}}

In fact, in ref. \cite{ana-miura}  we have constructed two solutions  for the  Miura-gauge transformation  (\ref{miura-gauge1}), namely,
\br
S_{\eps=1} = I +  s^{(-1)} =\begin{pmatrix}
1 & 0 \\
v & 1 
\end{pmatrix}
\qquad  \quad
S_{\eps=-1} = E^{(-1)} +  s^{(-2)} =\begin{pmatrix}
0 & \l^{-1} \\
1 & -\l^{-1}v 
\end{pmatrix} \label{jv1}
\er
where  $s^{(-i)} \in \lie_{-i}$ and   $E^{(-1)} =E_{-\a_1}+ \l^{-1}E_{\a_1}= \begin{pmatrix}0&\l^{-1}\\ 1&0 \label{em1}
\end{pmatrix}
$ corresponding to $\eps=+1$ and $\eps =-1$ respectively.

 {  {  Likewise  the $sl(2)$ case (\ref{jv1}), we now consider  $\lie = sl(3)$  where 
 \br
A_x^{mKdV} =\begin{pmatrix}
v_1 & 1 &0\\
0& -v _1+v_2&1\\
\l &0 & -v_2
\end{pmatrix}
\qquad  \quad
A_x^{KdV} =\begin{pmatrix}
0 & 1 &0\\
0&0 &1 \\
\l+J_1 & J_2 &0
\end{pmatrix} \label{axsl2}
\er
 $E \equiv E^{(1)} = E_{\a_1}+E_{\a_2} +\l E_{-\a_1-\a_2} =\begin{pmatrix}0&1&0\\ 0&0&1\\ \l &0&0 \label{em11}
\end{pmatrix}$ }} and 
propose  3 solutions for
$S$ satisfying (\ref{miura-gauge1}), 
\br
S_{1} = I +  s^{(-1)} +  s^{(-2)}=\begin{pmatrix}1&0&0\\ v_1&1&0 \\v_1^2-\pa_xv_1&v_2&1 \label{s1}
\end{pmatrix}
\er
leading to the Miura transformation
\br
J_1^{(1)} &=& -\pa_x (v_1+v_2) + v_1^2-v_1v_2+v_2^2, \nonu \\
J_2^{(1)} &=& v_1\pa_x (- 2 v_1 + v_2) + \pa_x^2v_1 +  v_1^2 v_2 - v_1 v_2^2 . \label{miura13}
\er
The second solution is given by
\br
S_{2} = E^{(-1)}  +  s^{(-2)} +  s^{(-3)}=\begin{pmatrix}0&0&\l^{-1}\\ 1&0&-\l^{-1}v_2 \\v_1-v_2&1&\l^{-1}(v_2^2+\pa_xv_2), \label{s2}
\end{pmatrix}
\er
for $E^{(-1)} = E^{\dagger}= E_{-\a_1} + E_{-\a_2} +  \l^{-1}E_{\a_1+\a_2} =\begin{pmatrix}0&0&\l^{-1}\\ 1&0&0\\ 0 &1&0 \label{em111}
\end{pmatrix}$, which leads  to the Miura transformation
\br
J_1^{(2)} &=& -\pa_x(v_1-2v_2) + v_1^2-v_1v_2+v_2^2, \nonu \\
J_2^{(2)} &=&  -v_2 \pa_x (v_1 + v_2) -\pa_x^2v_2 + v_1^2v_2 -v_1v_2^2.\label{miura23}
\er
A third solution is given by
\br
S_{3} = E^{(-2)}  +  s^{(-3)} +  s^{(-4)}=\begin{pmatrix}0&\l^{-1}&0 \\ 0&\l^{-1}(-v_1 + v_2)&\l^{-1} \\1&\l^{-1}A&-\l^{-1}v_1, \label{s3}
\end{pmatrix}
\er
for $E^{(-2)} = \l^{-1}E =\l^{-1}E_{\a_1} + \l^{-1}E_{\a_2} +  E_{-\a_1-\a_2}= \begin{pmatrix}0&\l^{-1}&0\\ 0&0&\l^{-1}\\ 1&0&0 \label{em2}
\end{pmatrix}$, and \newline
$\qquad A=\pa_x(v_1-v_2) +(v_1-v_2)^2\qquad $ leading  to the Miura transformation
\br
J_1^{(3)} &=& 2\pa_xv_1-\pa_xv_2 + v_1^2-v_1v_2+v_2^2, \nonu \\
J_2^{(3)} &=& (-v_1+ v_2) \pa_xv_1 + 2 (v_1 - v_2) \pa_xv_2 + \pa_x^2 (- v_1 + v_2) + v_1^2v_2 -v_1v_2^2. \label{miura33}
\er

In the appendix B we develop the example for $\lie =sl(4)$  by constructing 4  solutions for the Miura-gauge transformation  (\ref{miura-gauge1})  and notice a general pattern emerging that induces to the following  general proposition for  the $sl(n+1)$ case,
  \begin{itemize}
\item Given $E^{(1)}=\sum_{j=1}^{n}E_{\a_j} + \l E_{-(\a_1+\cdots +\a_n)}$, there are $ n $ generators of grade $q=$ $-1, -2, \ldots, -n$  commuting  with $  E^{(1)}$, namely, $\{ E^{(-1)},  E^{(-2)}, \cdots,  E^{(-n)} \}\in {\hat {\cal K}}$ (see  for instance  appendix of ref.\cite{olive-turok}).

\item Propose solution $S_i = E^{(-i)} + s^{(-i-1)} + s^{(-i-2)}  + \cdots + s^{(-i-n)}, \;\; i=0, 1, \ldots, n$ where $E^{(0)} = I$  and   $S_{i+n+1}=\l S_{i} $ such that the  algebraic structure  of gauge potentials (\ref{aa}) is preserved {\footnote{ It is clear that  $S$ contains a highest  grade  component commuting with $E^{(1)}$ in order to preserve the  common graded structure the two gauge potentials $A_{x}^{KdV}$ and $A_{x}^{mKdV}.$}}.

\item Each  solution generates  a Miura transformation $J_a^{(i)}, a=1, \ldots, n+1$.

\end{itemize}

{ {Notice that  the Miura transformation  $S$  acting on the zero curvature  representation  (\ref{zcc})   transforms  both gauge potentials, i.e.,  $A_{x }^{mKdV} \rightarrow A_{x }^{KdV}$    and  $A_{t_N}^{mKdV} \rightarrow A_{t_N }^{KdV}, $  for all values of $N$.  This implies that the  gauge-Miura transformation  $S$ is universal within the hierarchy in the sense that   all flows  (labeled by $N$)  are transformed by the same $S$.}}

%

\section{Equations of Motion}
We now  conjecture  that  the equations of motion of  the  two hierarchies  are connected as in the well  known case of the $A_2$  KdV and mKdV equations,
\br
4 \pa_t J - \pa_x^3 J -6 J \pa_x J = {\cal P}_{\eps}   \( 4 \pa_t v - \pa_x ( \pa_x^2 v - 2 v^3 ) \) = 0, \quad  {\cal P}_{\eps} = (  \eps \pa_x - 2 v ), \quad \eps = \pm1  \label{jv}
\er 
In fact, we argue  that  the equations of motion  of the generalized  $A_n$ mKdV  and KdV hierarchies are related    by a matrix  operator ${\cal P}$, as $ \[ KdV ( J_i)\] = {\cal P} \; \;  \[ mKdv  (v_i)\]$, or in components,
\br
 \begin{pmatrix} \pa_{t}J_1 \\ \pa_{t} J_2 \\ \vdots \\   \pa_{t} J_n \\  \end{pmatrix} = {\cal P} \begin{pmatrix} \pa_{t}v_1 \\ \pa_{t} v_2 \\ \vdots \\   \pa_{t_2} v_n \\ \end{pmatrix} \label{p}
\er
where the matrix operator  ${\cal P}$ is denoted  ${\cal P}= \( P_{ij}\)$.  

Explicitly  we have considered  the  equations of motion  (\ref{eqmot3}) and (\ref{eq-kdv-sl3}) for  $sl(3)$
written  in the form (\ref{p}) as
\br
 \begin{pmatrix} \pa_{t}J_1 \\ \pa_{t} J_2 \\  \end{pmatrix} = {\cal P} \begin{pmatrix} \pa_{t}v_1 \\ \pa_{t} v_2 \\ \end{pmatrix} \label{p3}
\er
It is clear that 
there is a different ${\cal P}$ operator associated to each solution of the Miura-Gauge transformation  (\ref{miura-gauge1}).

\begin{itemize}
\item
 For $S_1$  (\ref{s1}) 
 and Miura   (\ref{miura13}), 
\br
P_{11}^{(1)}&=& -\pa_x+2v_1-v_2 , \nonu \\
P_{12}^{(1)}&=& -\pa_x-v_1+2v_2 ,  \nonu \\
P_{21}^{(1)}&=& \pa_x^2-2v_1\pa_x-2\pa_xv_1+\pa_xv_2+2v_1v_2 \purple{-} v_2^2 , \nonu \\
P_{22}^{(1)}&=&  v_1\pa_x-2v_1v_2+v_1^2.
\er

\item Likewise for  the Miura-gauge transformation $S_2$ (\ref{s2})   and Miura (\ref{miura23}),
\br
P_{11}^{(2)} &=& -\pa_x+2v_1-v_2, \nonu \\
P_{12}^{(2)} &=& 2\pa_x-v_1+2v_2, \nonu \\
P_{21}^{(2)} &=& -v_2\pa_x+2 v_1 {v_2} -v_2^2, \nonu \\ 
P_{22}^{(2)} &=& -\pa_x^2 {- v_2 \pa_x} -\pa_xv_1-\pa_xv_2-2 v_1 {v_2} +v_1^2.
\er
 \item For $S_3$  given in (\ref{s3}) and Miura (\ref{miura33}), 
\br
P_{11}^{(3)} &=&  2 \pa_x + 2 v_1 - v_2 , \nonu \\
P_{12}^{(3)} &=& -\pa_x - v_1 + 2v_2,  \nonu \\
P_{21}^{(3)} &=& -\pa_x^2-v_1\pa_x {+}v_2\pa_x-\pa_xv_1+2\pa_xv_2+2v_1{v_2}-v_2^2,  \nonu \\
P_{22}^{(3)} &=& \pa_x^2+2v_1\pa_x-2v_2\pa_x {+} \pa_xv_1-2\pa_xv_2+v_1^2-2v_1{v_2}.
\er
\end{itemize}
For the $sl(4)$ case   the same procedure can be   employed   for each  associated Miura-gauge transformation given in appendix B.



\section{B\"acklund Transformation}

In this section we generalize  the results of ref.  \cite{ana-miura}  by constructing the B\"acklund transformation   for the $A_n$-KdV hierarchy  from the Miura   
and  B\"acklund-gauge transformations  constructed  for the $A_n$-mKdV hierarchy. 
Consider the B\"acklund-gauge transformation for the  $A_n$-mkdV hierarchy proposed in \cite{lobo},
\begin{equation}
U(\phi_i,\psi_i) A^{mKdV} _{\mu}({\phi_i}) = A^{mKdV} _{\mu}({\psi_i})U(\phi_i,\psi_i)  +\pa_{\mu}U(\phi_i,\psi_i) 
\label{b-mKdV}
\end{equation}
where $u_i=\pa_x\phi_i, \;\; v_i = \pa_x \psi_i$. Also, the Miura-gauge transformation can be written as,
\begin{equation}
A_{\mu}^{KdV}(J_i)= S A_{\mu}^{mKdV}(v_i)S^{-1} + S\pa_{\mu} S^{-1}.  \label{miura-gauge}
\end{equation}

Let ${K(I,J)}$ be the generator of  the B\"acklund-gauge transformation for the  $A_n$-KdV hierarchy, i.e.,
\begin{equation}
{ K(I_i,J_i) A^{KdV} _{\mu}(I_i) = A^{KdV} _{\mu}(J_i)K(I_i,J_i)  +\pa_{\mu}K(I_i,J_i) }
\label{b-KdV}
\end{equation}

Inserting  (\ref{miura-gauge}) into   (\ref{b-mKdV}) , we find
\begin{equation}
{K(I_i,J_i) =S(v_i) U(\phi_i,\psi_i)S^{-1}(u_i)} \label{b-g-kdv1}
\end{equation}

Notice that ${K(I_i,J_i)}$  given {   { by the  rhs of }} (\ref{b-g-kdv1}) depend upon  mKdV variables, $u_i$ and $v_i$.
In ref. \cite{ana-miura}  we  have shown that the gauge-B\"acklund transformation ${K(I_i,J_i)}$ was entirely written in terms of KdV variables, $I$ and $J$ if  we use  two different Miura-gauge transformations, for the right $S({u})=S_{+\eps}$ and for the  left $S({v})^{-1} = S_{-\eps}^{-1}$ multiplications (see (\ref{jv1})).
Here we will follow the same  principle by choosing  different  $S$ solutions for  right and left multiplication  in (\ref{b-g-kdv1}).

B\"acklund transformation for the $A_n$ Toda  theory was   first proposed  in \cite{Fordy-Gibb} and in \cite{lobo},  it was  generalized   as a gauge transformation constructed  and classified  according to  a graded  affine Lie algebraic  structure . Such construction was shown to be 
 universal within the  hierarchy, i.e.,  in the sense that it is  same  for all  positive and negative flows {\footnote{ Observe that the relativistic Toda model  correspond to the first negative flow   \cite{aratyn}}}.


Consider, {  {as an illustration, }}  the type I  B\"acklund-gauge transformation for $sl(3)$ mKdV hierarchy  (see for instance \cite{lobo}),

\begin{equation}
U (\phi_i,\psi_i)= 
\begin{pmatrix}
1 & 0 & \frac{\sigma}{\lambda} e^{-\phi_2-\psi_1} \\
\sigma e^{\phi_1+\psi_1-\psi_2} & 1 & 0 \\
0 & \sigma e^{-\phi_1+\phi_2+\psi_2} & 1
\end{pmatrix} 
\label{6.5}
\end{equation}
which  yield  {   { from (\ref{b-mKdV})  for $ A_x$ }} the following  equations corresponding to  the mKdV  B\"acklund transformation,
\br
u_1-v_1 = \s \(e^{\phi_1+\psi_1-\psi_2}-e^{-\phi_2-\psi_1}\), \qquad u_2-v_2 = \s \(e^{-\phi_1+\phi_2+\psi_2}-e^{-\phi_2-\psi_1}\) \label{bk-mkdv}
\er
where $u_i=\pa_x\phi_i, \;\; v_i = \pa_x \psi_i, \; i=1,2$ 
and  take  the   two solutions ${S_1(u_i)}$ and ${S_3(v_i)}$  given by (\ref{s1}) and (\ref{s3}) 
to  be inserted as  right and left multiplications in (\ref{b-g-kdv1})  yielding the following matrix elements,
\begin{eqnarray}
K_{11}&=& (-u_1+\s e^{\phi_1+\psi_1-\psi_2} \label{11})\lambda^{-1} \label{Backlund-001}  \\ 
K_{22}&=& (-u_2-v_1+v_2 +\s e^{-\phi_1+\phi_2+\psi_2}\label{22})\lambda^{-1} \label{Backlund-002}\\
K_{33}&=& (-v_1+\s e^{-\phi_2-\psi_1} \label{33})\lambda^{-1} \label{Backlund-003}\\
K_{12} &=& K_{23}= \lambda^{-1}  \\
K_{31} &=& 1 + Y \lambda^{-1} \label{Backlund-005}
\end{eqnarray}

\begin{align}
K_{13}&= 0\\
K_{21}&=\(\pa_xu_1-u_1^2+u_1u_2 -u_1\s e^{-\phi_1+\phi_2+\psi_2} -(v_1-v_2)(-u_1 + \s  e^{\phi_1+\psi_1-\psi_2})\)\lambda^{-1}  \label{21}\\
K_{32}&= \(\pa_xv_1 -\pa_xv_2 + v_1^2 + v_2^2 -2v_1v_2 + u_2v_1-u_2\s e^{-\phi_2-\psi_1}-v_1\s e^{-\phi_1+\phi_2+\psi_2}\)\lambda^{-1}\label{32}
\end{align}
where
\begin{equation}
\begin{split}
Y = & u_1v_1 \s e^{-\phi_1+\phi_2+\psi_2} + (\pa_xu_1 - u_1^2 +u_1u_2)(-v_1 + \s e^{-\phi_2-\psi_1})  \\
&\quad +(\pa_xv_1-\pa_xv_2 + v_1^2+v_2^2-2v_1v_2)(-u_1 + \s  e^{\phi_1+\psi_1-\psi_2})  \\
\end{split}
\end{equation}
We now show how to  re-write  the B\"acklund matrix $K(I_i,J_i)$ in  terms of KdV variables. 
Subtracting the diagonal terms and using (\ref{bk-mkdv}),
\begin{align}
K_{11}-K_{22}&= \(-u_1+v_1 +u_2-v_2+\s e^{\phi_1+\psi_1-\psi_2}   -\s e^{-\phi_1+\phi_2+\psi_2}\)\lambda^{-1} =0\\
K_{22}-K_{33}&= \(-u_2+v_2+  \s e^{-\phi_1+\phi_2+\psi_2} - \s e^{-\phi_2-\psi_1} \)\lambda^{-1} =0
\end{align}
and henceforth
\begin{equation}
K_{11}=K_{22}=K_{33}\equiv {{1}\over {3}} Q \lambda^{-1}.
\label{Q}
\end{equation}
Acting with $x$ derivative on $Q$ and re-arranging terms,
\begin{equation*}
\begin{split}
\pa_x Q &= \lambda \pa_x (K_{11}+K_{22}+K_{33}) \\
&=\pa_x (-u_1-u_2-2v_1+v_2) +  u_1\sigma(e^{-\phi_1 + \psi_1 - \psi_2} - e^{-\phi_1 +\phi_2 + \psi_2})\\
&\quad+ u_2\sigma(e^{-\phi_1+\phi_2 + \psi_2} - e^{-\phi_2 -\psi_1}) +  v_1\sigma(e^{\phi_1 + \psi_1 - \psi_2} - e^{-\phi_2 -\psi_1}) +  v_2\sigma(e^{-\phi_1+\phi_2 + \psi_2} - e^{\phi_1 + \psi_1 - \psi_2}).
\end{split}
\end{equation*}

After eliminating  the exponentials  from the mKdV B\"acklund transformations  (\ref{bk-mkdv}) and subsequent use of   Miura transformations  (\ref{miura13}) and (\ref{miura33}), i.e.,
\begin{eqnarray}
I_1(u_i) &=& J_1^{(1)}(u_i)= -\pa_xu_1-\pa_xu_2 +  u_1^2+u_2^2-u_1u_2\\
J_1(v_i) &=& J_1^{(3)}(v_i)=2\pa_xv_1-\pa_xv_2  + v_1^2+v_2^2-v_1v_2 \label{IJ}
\end{eqnarray}
we find,
\begin{equation}
\pa_xQ = I_1 -  J_1.
\end{equation}

The  $K_{21}$ matrix element also can be  written enterely  in terms of KdV variables, e.g., replacing the $\partial_xu_1$ term using \eqref{Backlund-001},
\begin{equation*}
\begin{split}
\lambda K_{21}&=-\frac{1}{3}\partial_xQ + u_1\s (e^{\phi_1+\psi_1-\psi_2}-e^{-\phi_1+\phi_2+\psi_2})-u_1(u_1-u_2-v_1+v_2)   \\
&= -{{1}\over {3}} \pa_x Q
\end{split}
\end{equation*}
where  we have used (\ref{bk-mkdv}).  Likewise,  using the Miura transformation (\ref{IJ})  in \eqref{32} and then replacing the remaining $\partial_xv_1$ using \eqref{33} we find,
\begin{equation*}
\begin{split}
\lambda\,K_{32}&= J_1 -\pa_xv_1-v_1v_2 + u_2v_1-u_2\s e^{-\phi_2-\psi_1}-v_1\s e^{-\phi_1+\phi_2+\psi_2}\\
&=J_1 + \frac{1}{3}\partial_xQ + v_1\s (e^{-\phi_2 - \psi_1} - e^{-\phi_1 + \phi_2 + \psi_2}) + v_2(u_2 - v_2) = J_1 + {{1}\over {3}}\pa_xQ
\end{split}
\end{equation*}
where  we have used (\ref{bk-mkdv}) to eliminate the exponentials. Finally doing the same procedure for    \eqref{Backlund-005} we find
\br
K_{31}= 1 +{{Y}\over {\lambda}}
\er
where $Y =\s^3-{{Q}\over {3}}^3 + {{1}\over {3}}Q J_1$ and 
\begin{equation}
K (I,J)= 
\begin{pmatrix}
{{1}\over {3 \lambda }}Q & {{1}\over {\lambda}} & 0 \\[8pt]
-{{1}\over {3\lambda} }(I_1-J_1) &  {{1}\over {3\lambda}}Q & {{1}\over {\lambda}}  \\[8pt]
1 +{{Y}\over {\lambda}} & {{1}\over {3\lambda}}(I_1+2J_1) & {{1}\over {3\lambda}}Q 
\end{pmatrix}. \label{k}
\end{equation}
It can be verified that $det \;K= {{1}\over {\lambda}}  +{{\s^3}\over {\lambda^2}}$ .

We have verified that the very same argument follows 
if we instead of the pair $S_3(v_i)$ and $S_1^{-1}(u_i)$,  use in (\ref{b-g-kdv1}) $S_1(v_i)$ and $S_2^{-1}(u_i)$.  The resulting B\"acklund-gauge  transformation $K$  for the KdV hierarchy has the same form as (\ref{k}) but now written in terms of the  corresponding Miura  fields given in (\ref{miura13}) and (\ref{miura23}) and  
so  is  the resulting  B\"acklund-gauge generator  for the remaining pair $S_2, S_3$.

Let us now  write down explicit B\"acklund transformation for the $A_2$-mKdV  system {  {  (\ref{j2}) and (\ref{eq-kdv-sl3})   according to $t=t_2$}}. Employing the  gauge-B\"acklund transformation \eqref{k}  in \eqref{b-KdV} where  gauge potentials  $A_{\mu}^{KdV}$ are given by,
\begin{eqnarray}
A_x^{KdV}=
\begin{pmatrix}
	0 & 1 &0 \\[6pt]
	0 & 0 & 1 \\[6pt]
	\l+J_2 & J_1 & 0
\end{pmatrix}, \quad
A_t^{KdV}=
\begin{pmatrix}
	-{{2J_1}\over {3}} & 0 &1 \\[6pt]
	\l+{{2}\over{3}}\pa_xJ_1+J_2 & {{J_1}\over {3}}& 0 \\[6pt]
	-{{2}\over {3}}\pa_x^2J_1-\pa_xJ_2 & \l+{{1}\over {3}}\pa_xJ_1+J_2 & {{J_1}\over {3}}
\end{pmatrix} \label{at2} .
\end{eqnarray}
The nontrivial equations obtained  from  (\ref{b-KdV}) for $A_x^{KdV}$  are
\begin{eqnarray} 
I_2-J_2   &=& \pa_xJ_1 - {{1}\over {3}} (I_1-J_1)Q\label{31x} \\
\pa_x(I_1-J_1)  &=&- 3I_2 +J_1Q  -{{1}\over {9}}Q^3 +3\s^3  \label{21x}\\
\pa_x(I_1+2J_1) &=&-3 J_2 +I_1Q  -{{1}\over {9}}Q^3 +3\s^3\label{32x}
\end{eqnarray}
which are compatible in the sense that a linear combination  of  any two yields the third.
Those correspond to matrix elements $(3,1), (2,1)$ and $(3,2)$ respectively.

As for the  $A_t^{KdV}$ potential, the nontrivial eqns. corresponding to matrix diagonal elements are,
\begin{eqnarray}
\pa_tQ &=& \,\,2\pa_xI_1+3I_2-I_1Q+{{1}\over {3}}Q(I_1-J_1) +{{1}\over {9}}Q^3-3\s^3, \label{11t}\\
\pa_tQ &=&  \pa_x(I_1-2J_1)+3(I_2-J_2) + {{1}\over {3}}(I_1-J_1)Q, \label{22t}  \\
\pa_tQ &=& -\pa_xJ_1-3J_2 +J_1Q + {{1}\over {3}}Q(I_1-J_1) -{{1}\over {9}}Q^3+3\s^3. \label{33t}
\end{eqnarray}

Other nontrivial eqns. are,
\begin{eqnarray}
3\pa_t(I_1-J_1)&=&6\pa_x^2I_1 +9\pa_xI_2 -2I_1^2+J_1^2+ I_1J_1-\(2\pa_x(I_1-J_1)+3(I_2-J_2)\)Q, \label{21t} \\
3\pa_t(I_1+2J_1)&=&6\pa_x^2J_1 +9\pa_xJ_2 +I_1^2-2J_1^2 + I_1J_1 +\(\pa_x(I_1-J_1)+3(I_2-J_2)\)Q, \label{32t}
\end{eqnarray}

\begin{equation}
\begin{split}   
\partial_t\(- \frac{Q^3}{27} + \frac{1}{3}QJ_1\) &= \frac{1}{9}\left(J_1 \left(\partial_x(4I_1-J_1)+6 I_2-3 J_2-2 I_1 Q+\frac{Q^3}{9}-3 \sigma ^3\right) \right.\\ & \left. \quad+I_1 \left(\partial_x(2I_1+J_1)+3 I_2+3 J_2+\frac{2 Q^3}{9}-6 \sigma ^3\right) \right.\\ & \left.\quad +Q \partial_x\left(-3 I_2+3J_2-2\partial_x I_1+2 \partial_xJ_1\right)-J_1^2 Q\right)
\end{split}
\label{n14-005}
\end{equation}
and correspond to matrix elements $(2,1)$, $(3,2)$ and $(3,1)$ respectively. 

Subtracting (\ref{11t}) -   (\ref{22t}) and $ (\ref{22t})-(\ref{33t})$ we eliminate $\pa_tQ$ and recover (\ref{32x}) and (\ref{21x}) respectively.
Moreover, substituting  (\ref{31x}) in  (\ref{22t}) we find
\br
\pa_t Q = \pa_x(I_1-J_1) + 2(I_2-J_2), \label{635}
\er
which leads directly to the eqns. of motion (\ref{j2})  for $I_1$ and $J_1$ by acting with $\pa_x$ in (\ref{635}), i.e.,
\br
\pa_t (I_1-J_1)= \pa_x^2(I_1-J_1) + 2\pa_x(I_2-J_2).
\er

Subtracting (\ref{32t}) and  (\ref{21t})  we find
\begin{equation}
3\pa_tJ_1=-2\pa_x^2(I_1-J_1)-3\pa_x(I_2-J_2) +(I_1^2-J_1^2)+ (\pa_x(I_1-J_1)+2(I_2-J_2))Q\label{j1t}
\end{equation}
Acting with $\pa_t$  on eqn. (\ref{31x}) and using  the fact that $\pa_x Q=I_1-J_1$, 
\begin{equation*}
\begin{split}
3\pa_t(I_2-J_2) &= \pa_x \(3\pa_t J_1 -Q\pa_t Q\) \\
&=\pa_x \(Q (-(I_2-J_2) +  \pa_xJ_1-{{1}\over {3}}(I_1-J_1)Q )\) \\
&+\pa_x \(-2\pa_x^2(I_1-J_1)-3\pa_x(I_2-J_2) +  (I_1^2-J_1^2)\)
\end{split}
\end{equation*}
where we  have used (\ref{j1t}) and  substituted $\pa_tQ$ using (\ref{22t}).  After  making use of (\ref{31x})  to eliminate the first term (proportional to $Q$) we end up with  the equation of motion for $I_2$ and $J_2$ (\ref{eq-kdv-sl3}), i.e.,
\begin{equation}
\pa_t(I_2-J_2) =-\frac{1}{3}\pa_x \(2\pa_x^2(I_1-J_1)  + 3\pa_x(I_2-J_2)-(I_1^2-J_1^2)\)
\end{equation}


{  {Notice that, since the B\"acklund transformation  generator $K (I,J)$  in (\ref{k}) is the same  for all flows,   the same procedure can be  employed  for higher flows $t=t_{N}$ generating therefore,  the B\"acklund transformation   for the entire hierarchy in a systematic manner.    The same  can be extended to  a general $A_n$  KdV/mKdV integrable models }}.


\section{Discussion and Further Developments}

The  gauge invariance of  zero curvature representation was explored in order to map the $A_n$-mKdV hierarchy into its counterpart, the $A_n$-KdV hierarchy.  Such map is known as {\it generalized Miura-gauge transformation} and is generated by a gauge transformation denoted by $S$.  We have shown  by developing explicit examples,  that  $S$ has the virtue of preserving the algebraic structure  of the Lax operators (i.e., two dimensional gauge potentials $A_{\mu}^{mKdV}$ and $A_{\mu}^{KdV}$).  An interesting  discovery is that $S$ is not unique, as it was already been suggested from the well known $sl(2)$ example (\ref{jv}) in which  the  Miura transformation is two-fold degenerated,  parameterized by $\eps = \pm1$.  The $A_n$-Miura-gauge transformation   present  a $n+1$ degeneracy and  was shown to be classified  according to the elements of the Kernel of $E^{(1)}, {{\hat {\cal K}}}$, supplemented  with the  identity element $I$.  The role of the subgroup of transformations  generated by  such subset of generators  is still not entirely clear and is currently under investigation.

The B\"acklund transformation for the  KdV hierarchy, in turn was inherited  from the gauge-B\"acklund transformation  for the mKdV hierarchy by  left-right multiplication  by Miura-gauge generators as in (\ref{b-g-kdv1}), i.e., $K(J,I) = S(I) U(\phi,\psi)S^{-1}(J) $.

  A surprising feature of the construction, is a nontrivial change of mKdV to KdV variables $(\phi,\psi) \rightarrow (J,I)$ which was shown to be  possible    if the left and right  Miura  transformations appearing in (\ref{b-g-kdv1})  correspond to different degenerate solutions associated to ${\hat {\cal K}}$.  Such fact  was already realized  for the $sl(2)$ case in \cite{ana-miura} and was explicitly verified  for several combinations  of Miura solutions for $sl(3)$ example.

As a matter of fact, the resulting gauge B\"acklund transformation  was  entirely  written in terms of KdV variables and appears as an  {\it universal} object  within the hierarchy, in the sense  that it is the same for all flows.  As a byproduct, it generates, in a systematic manner, the B\"acklund transformation  for all flows.  Moreover, this method  provides a classification of integrable defects  as proposed in \cite{corrigan}  and may also  be    extended  to  other    generalized  type II B\"acklund transformations from mKdV to KdV hierarchies \cite{lobo}.   The framework employed  in this paper may be  also extended to  Lie algebras other than  $A_n$, whose Dynkin diagram may connect  more than two nearest neighbours, e.g., $B_4, E_6,E_7,E_8$ in the line of ref. \cite{bristow} or to non-simply laced algebras.  Application to twisted  algebras may also  provide  interesting examples in the lines  of  ref.  \cite{robertson}.


\vspace{5mm}
{\bf Acknowledgments}
JFG and AHZ thank CNPq and FAPESP for partial  support. JMCF and GVL thank CNPq and Capes  respectively for  financial support.

%
\appendix

\section{ Affine Algebraic Structure}

Consider  an  affine Kac-Moody algebra   $\hat \lie$ defined by 
\br
\lb H_i^{(l)}, H_j^{(k)}\rb &=& \kappa l \d_{l+k,0} \d_{i,j}, \quad \nonu \\
\lb H_i^{(l)}, E_{\a}^{(k)}\rb &=&\a^{i}E_{\a}^{(l+k)}\nonu \\
\lb E_{\b}^{(l)}, E_{\a}^{(k)}\rb& =&
\begin{cases}
	\eps (\a, \b)E_{\a+\b}^{(l+k)}, \quad \a+\b = {\rm root},\nonu \\
	{{2}\over {\a^2}}\a \cdot H^{(l+k)} + \kappa l  \d_{l+k,0}, \quad \a+\b =0,\nonu \\
	0\quad  {\rm otherwise}.
\end{cases}
\label{a1}
\er
{   { $ i,j=1,\cdots rank\;\;  \lie,  \quad l,k \in Z$.}}
Let   $Q$ be a grading operator and  consider a decomposition of  the  affine algebra $\hat \lie$ into grades subspaces, $\lie_a$ such that,
\br 
\hat \lie = \sum_{a\in Z} \lie_a, \qquad \lb Q, \lie_a\rb = a \; \lie_a, \qquad \lb \lie_a, \lie_b\rb \in \lie_{a+b}
\label{a2}
\er
In this paper we  shall discuss  $\hat \lie = \hat {sl}(n+1)$ endowed  with  the  principal gradation in which
\br Q^{ppal} = (n+1)d + \sum_{a=1}^{n} \mu_a\cdot h
\label{a3}
\er 
where  $d$ is the derivation operator, i.e.,
\br
\lb d, T_i^{(l)}\rb = l  T_i^{(l)}, \qquad   T_i^{(l)} =  h_i^{(l)} \; {\rm or }\;\; E_{\a}^{(l)}
\er
and
\br
\lb \mu_a \cdot h^{(l)}, E_{\a}^{(k)}\rb = ({\mu_a \cdot \a }) E_{\a}^{(l+k)}.
\er
Here $\mu_a$  and $\a_a$ are  the fundamental weights and  simple roots respectively,
${\mu_a \cdot \a_b} = \d_{a,b}$, \quad $a,b=1,\cdots, n$, and  have normalized all roots  of $\hat sl(n+1)$ such that $\a^2=2$. 
The  operator $Q$ in (\ref{a3})  induces the following graded subspaces,
\br
\lie_{l(n+1)} &=&\{ h_1^{(l)}, \cdots ,  h_n^{(l)}\},\nonu \\
\lie_{l(n+1) +1} &=&\{ E_{\a_1}^{(l)}, \cdots ,  E_{\a{_n}}^{(l)}, E_{-\a_1 - \cdots -\a_n}^{(l+1)}\},\nonu \\
\lie_{l(n+1)+2} &=&\{ E_{\a_1+\a_2}^{(l)}, E_{\a_2+\a_3}^{(l)}, \cdots ,  E_{\a_{n-1}+\a{_n}}^{(l)}, E_{-\a_1 - \cdots -\a_{n-1}}^{(l+1)},   E_{-\a_2 - \cdots -\a_{n}}^{(l+1)}\},\label{a4} \\
&\vdots&  \nonu \\
\lie_{l(n+1) +n }&=&\{ E_{-\a_1}^{(l+1)}, \cdots ,  E_{-\a{_n}}^{(l+1)}, E_{\a_1 +\cdots + \a_n}^{(l)}\}. \nonu
\er
where $h_i^{(l)} = \a_i\cdot H^{(l)}$.


  Let ${\hat {\cal {K}}}$ denote the Kernel of $E$ composed  by all elements within each subspace  of grade $q$ commuting with $E$ {\footnote{In fact there are precisely $n$ generators commuting with $E$  of grade  given by the  exponents   modulo the Coxeter number
  $h$ which in the  $A_n$  case are $q= 1, \cdots ,n$ modulo $n+1$}.
  In particular denote them by $E^{(-q)}\in  {\hat {\cal {K}}}, q=1, \cdots n$  mod $(n+1)$,  (see appendix of ref. \cite{olive-turok}) e.g.,
\br
E^{(-n)} &=&  \sum_{k=1}^{n}E_{\a_k}^{(-1)}  + E_{-(\a_1+\cdots + \a_n)}^{(0)} ,\nonu \\
E^{(-n+1)} &=&\sum_{k=1}^{n-1} E^{(-1)}_{\alpha_k+\alpha_{k+1}}
+E^{(0)}_{-(\alpha_1 + \cdots + \alpha_{n-1})}+
E^{(0)}_{-(\alpha_2 + \cdots+\alpha_{n})}, \nonu \\
 & \vdots &  \nonu \\
E^{(-1)} &=& \sum_{k=1}^{n}E_{-\a_k}^{(0)}  + E^{(-1)}_{-(\a_1+\cdots + \a_n)}
\er


\section{Miura Transformation for $sl(4)$ -  Example}

 We now consider  $\lie = sl(4)$  where 
$E \equiv E^{(1)} = E_{\a_1}+E_{\a_2} + E_{\a_3}+ \l E_{-\a_1-\a_2-\a_3}$  and  propose  4 solutions for $S$ satisfying (\ref{miura-gauge1}) { {acting on 
 \br
A_x^{mKdV} =\begin{pmatrix}
v_1 & 1 &0&0\\
0& -v _1+v_2&1&0\\
0&0&-v_2+v_3 &0 \\
\l &0 &0 & -v_3
\end{pmatrix}
\qquad  \quad
A_x^{KdV} =\begin{pmatrix}
0 & 1 &0&0\\
0&0 &1&0 \\
0&0 &0&1 \\
\l+J_1 & J_2 &J_3 &0
\end{pmatrix} \label{axsl4}
\er
}}
Let 
\br
S_{1} &=& I +  s^{(-1)} +  s^{(-2)} + s^{(-3)}\nonu \\
	   &=&\begin{pmatrix}
							1&0&0&0 \\ 
							v_1&1&0&0 \\
							-\pa_xv_1 + v_1^2&v_2&1 &0 \\ 
							\pa_x^2v_1 -3v_1\pa_xv_1 + v_1^3& -\pa_x v_1 -\pa_xv_2 + v_1^2 - v_1 v_2+v_2^2& v_3&1\label{s11}
\end{pmatrix} \label{s11}
\er
leading to the Miura transformation
\br
J_1^{(1)}&=&-\pa_x(v_1+v_2+v_3) + v_1^2 +v_2^2 + v_3^2 - v_1 v_2 - v_2v_3, \nonu \\
J_2^{(1)}&=& \pa_x^2 (2 v_1 + v_2) + (-4v_1+v_2) (\pa_x v_1) +2(v_1-v_2) (\pa_x v_2) + v_2 (\pa_x v_3) + v_1^2 v_2 - v_1 v_2^2 + v_2^2 v_3 - v_2 v_3^2, \nonu \\ 
J_3^{(1)}&=&  -\pa_x^3v_1 +v_1 \pa_x^2 ( 2 v_1 - v_2 ) + \pa_x (2 v_1 - v_2 - v_3) (\pa_x v_1) + (v_2^2 +v_3^2 - 2v_1v_2-v_2v_3) (\pa_xv_1)  \nonu \\
&+& (-v_1^2+v_1v_2)(\pa_x v_2) + (v_1^2-v_1v_2)(\pa_xv_3) + v_1^2v_2v_3-v_1v_2^2v_3+v_1v_2v_3^2-v_1^2v_3^2.
  \label{miura1}
\er
The second solution is given by
\br
S_{2} &=& E^{(-1)}  +  s^{(-2)} +  s^{(-3)} +s^{(-4)} \nonu \\
          &=&\begin{pmatrix}	0&0&0&\l^{-1}\\ 
							1&0&0& {-} \l^{-1} v_3\\
							v_1-v_3&1&0& \l^{-1}(\pa_xv_3 + v_3^2)\\  
							-\pa_xv_1 +2\pa_xv_3 + v_1^2-v_1v_3+v_3^2& v_2-v_3&1& {-} \l^{-1}(\pa_x^2v_3 + 3v_3\pa_xv_3+ v_3^2), \label{s22}
\end{pmatrix}  \label{s22}
\er
for $E^{(-1)} = E^{\dagger}= E_{-\a_1} + E_{-\a_2} +  E_{-\a_3}+\l^{-1}E_{\a_1+\a_2+\a_3} $, which leads  to the Miura transformation
\br
J_1^{(2)}&=& -\pa_x(v_1+v_2-3v_3) +v_1^2+v_2^2+v_3^2 - v_1v_2 - v_2v_3, \nonu \\
J_2^{(2)}&=& \pa_x^2 (v_1 - 3 v_3) - 2 v_1 (\pa_x v_1) + (v_1-v_3) (\pa_xv_2) - 2 v_3 (\pa_xv_3) + v_1^2v_2-v_1v_2^2+v_2^2v_3 -v_2v_3^2, \nonu \\
J_3^{(2)}&=& \pa_x^3 v_3 +v_3 \pa_x^2 (v_1 + v_3) + \pa_x (v_1 + v_2) (\pa_xv_3) + (v_3^2 - 2 v_1 v_3) (\pa_x v_1) + v_1v_3 (\pa_xv_2) \nonu \\
&+& (-v_1^2-v_2^2+v_1v_2+v_2v_3)(\pa_xv_3) + v_1^2v_2v_3-v_1v_2^2v_3+v_1v_2v_3^2-v_1^2v_3^2.
\label{miura2}
\er
A third solution is given by
\br
S_{3} = E^{(-2)}  +  s^{(-3)} +  s^{(-4)} +  s^{(-5)}=\begin{pmatrix}0&0&\l^{-1}&0 \\ 0&0&\l^{-1}(v_3-v_2)&\l^{-1} \\1&0&\l^{-1}s_{33}&-\l^{-1}v_2\\
v_1-v_2&1&\l^{-1}s_{43} &\l^{-1}s_{44} \label{s33}
\end{pmatrix}
\er
for $E^{(-2)} = \l^{-1}E =\l^{-1}E_{\a_1+\a_2} + \l^{-1}E_{\a_2+\a_3} +  E_{-\a_1-\a_2}  +  E_{-\a_2-\a_3}$  and 
\br
s_{33} &=& \pa_x (v_2 - v_3) + (v_2 - v_3)^2, \qquad s_{44} = \pa_x (2 v_2 - v_3) + (v_2 - v_3)^2, \nonu \\
 s_{43} &=& \pa_x^2(- v_2 + v_3) - 3( v_2 - v_3) \pa_x (v_2 - v_3) - (v_2 - v_3)^3, \nonu
\er 
leading  to the Miura transformation
\br
J_1^{(3)} &=& \pa_x(3 v_1 - v_2 - v_3) +v_1^2 + v_2^2 + v_3^2 - v_1 v_2 - v_2 v_3,  \nonu \\
J_2^{(3)} &=& \pa_x^2(- v_2 +2 v_3) - v_2 (\pa_x v_1) + 2(-v_2 + v_3)(\pa_x v_2) + (3 v_2 - 4v_3)(\pa_x v_3) + v_1^2 v_2 -v_1 v_2^2+v_2^2 v_3 -v_2 v_3^2, \nonu \\
J_3^{(3)} &=& -\pa_x^3 (v_2 - v_3) + (v_2 - v_3) \pa_x^2 (v_2 - 2 v_3) + \pa_x (v_1 - 2 v_3) \pa_x (v_2 - v_3) + (v_3^2 - 2 v_2 v_3) (\pa_x v_1) \nonu \\
&+& (-v_1^2 + v_1 v_2 - v_2 v_3) (\pa_x v_2) + (v_1^2 - v_2^2 -v_1 v_2 +2 v_2 v_3) (\pa_x v_3) + v_1^2v_2v_3-v_1v_2^2v_3+v_1v_2v_3^2-v_1^2v_3^2. \nonu \\
\label{miura3}
\er
The fourth solution is 
\br
S_{4} = E^{(-3)}  +  s^{(-4)} +  s^{(-5)} +  s^{(-6)}=\begin{pmatrix}0&\l^{-1}&0&0 \\ 0&\l^{-1}(v_2-v_1)&\l^{-1} &0\\0&\l^{-1}t_{32}&-\l^{-1}(v_3-v_1)&\l^{-1} \\1 &\l^{-1}t_{42}&
\l^{-1}t_{43} &-\l^{-1}v_1 \label{s4}
\end{pmatrix}
\er
for $E^{(-3)}  =\l^{-1}E_{\a_1} + \l^{-1}E_{\a_2} + \l^{-1} E_{\a_3} +  E_{-\a_1-\a_2-\a_3}$  and 
\br
t_{32} &=& \pa_x (v_1 - v_2) + (v_1 - v_2)^2, \nonu  \\
t_{42} &=& \pa_x^2 (-v_1 + v_2) - 3 (v_1 - v_2) \pa_x (v_1 - v_2) + (v_1-v_2)^3, \nonu \\
t_{43} &=& \pa_x (2 v_1 - v_2 - v_3) +  v_1^2 + v_2^2 + v_3^2 - v_1 v_2 - v_1 v_3 - v_2  v_3, \nonu \\
\er 
leading  to the Miura transformation
\br
J_1^{(4)} &=& \pa_x ( 3 v_1 - v_2 - v_3) + v_1^2+v_2^2+v_3^2 - v_1 v_2 - v_2 v_3, \nonu \\
J_2^{(4)} &=&  \pa_x^2 (-3 v_1 + 2 v_2 +v_3) + 2(-v_1 + v_2) (\pa_x v_1) + (3 v_1 - 4 v_2 + v_3)(\pa_x v_2) + 2(v_2 - v_3) (\pa_x v_3)  \nonu \\
&+&  v_1^2 v_2 - v_1 v_2^2 + v_2^2 v_3 - v_2 v_3^2, \nonu \\
J_3^{(4)} &=&  \pa_x^3 (v_1 - v_2) + (v_1 - v_2) \pa_x^2 (v_1 - 2 v_2 + v_3) - \pa_(v_1 - v_2)\pa_x (2 v_2 - v_3) + (- v_3^2 - v_1 v_2 + v_2 v_3)(\pa_x v_1) \nonu \\ 
&+& (- v_1^2 + v_3^2 + 2 v_1 v_2 - 2 v_2 v_3 + v_1 v_3)(\pa_x v_2) + (v_1^2 -v_2^2 - 2 v_1 v_3 + 2 v_2 v_3)(\pa_x v_3)   \nonu \\
&+& v_1^2v_2v_3-v_1v_2^2v_3+v_1v_2v_3^2-v_1^2v_3^2.
\label{miura3}
\er

\end{document}